\documentclass[twocolumn]{aastex63}
\usepackage{inputenc}
\usepackage{comment}
\usepackage{colortbl}
\usepackage{graphicx}
\usepackage{soul}
\usepackage{hyperref}
\usepackage{multirow}

\received{99 Doc 2099}
\revised{99 Doc 2099}
\accepted{99 Doc 2099}
\submitjournal{ApJ}

\shorttitle{J0343}
\shortauthors{DiKerby et al. 2025}
\graphicspath{{./}{figures/}}
\begin{document}


\title{Discovery of a Pulsar Wind Nebula Candidate Associated with the Galactic PeVatron 1LHAASO J0343+5254u}

\author[0000-0003-2633-2196]{Stephen DiKerby}
\affiliation{Department of Physics and Astronomy \\
 Michigan State University,
East Lansing, MI 48820, USA}

\author[0000-0002-2967-790X]{Shuo Zhang}
\affiliation{Department of Physics and Astronomy \\
Michigan State University, East Lansing, MI 48820, USA}

\author[0000-0003-2423-4656]{T\"ul\"un Ergin}
\affiliation{Department of Physics and Astronomy \\
 Michigan State University, East Lansing, MI 48820, USA}

\author[0000-0001-7209-9204]{Naomi Tsuji}
\affiliation{Faculty of Science, Kanagawa University, 3-27-1 Rokukakubashi, Kanagawa-ku, Yokohama-shi, Kanagawa 221-8686}

\author[0000-0002-9709-5389]{Kaya Mori}
\affiliation{Columbia Astrophysics Laboratory, Columbia University, New York, NY 10027, USA}


\author[0000-0002-6606-2816]{Fabio Acero}
\affiliation{FSLAC IRL 2009, CNRS/IAC, La Laguna, Tenerife, Spain}   
\affiliation{AIM, CEA, CNRS, Universit\'e Paris-Saclay, Universit\'e de Paris, F-91191 Gif sur Yvette, France}


\author[0000-0001-6189-7665]{Samar Safi-Harb}
\affiliation{Department of Physics and Astronomy \\
University of Manitoba, Winnipeg, MB, R3T 2N2, Canada}

\author[0000-0001-8147-6817]{Shunya Takekawa}
\affiliation{Department of Applied Physics, Faculty of Engineering, Kanagawa University, 3-27-1 Rokkakubashi, Kanagawa-ku, Yokohama, Kanagawa 221-8686, Japan}

\author[0009-0001-6471-1405]{Jooyun Woo}
\affiliation{Columbia Astrophysics Laboratory, Columbia University, New York, NY 10027, USA}

\begin{abstract}

The astronomical origin of the most energetic galactic cosmic rays and gamma rays is still uncertain. X-ray followup of candidate ``PeVatrons'', systems producing cosmic rays with energies exceeding $1 \:\rm{PeV}$, can constrain their spatial origin, identify likely counterparts, and test particle emission models. Using $\sim 120 \:\rm{ks}$ of XMM-\textit{Newton} observations, we report the discovery of a candidate pulsar wind nebula, a possible counterpart for the LHAASO PeVatron J0343+5254u. This extended source has a power law X-ray spectrum with spectral index $\Gamma_X = 1.9$ - softer at greater distance from the center - and asymmetric spatial extension out to $\approx 2 \arcmin$. We conduct leptonic modeling of the X-ray and gamma ray radiation from this complex system, showing that a fully leptonic model with elevated IR photon fields can explain the multiwavelength emission from this source, similar to other VHE pulsar wind nebulae; excess gamma ray emissivity not explained by a leptonic model may be due to hadronic interactions in nearby molecular cloud regions, which might also produce detectable astroparticle flux.
    
\end{abstract}


\section{Introduction and Context}
\label{sec:Intro}

Galactic PeVatrons, astronomical systems producing cosmic rays (CRs) with energies reaching or exceeding $1 \: \rm{PeV} = 10^{15} \:\rm{eV}$, are the most energetic astrophysical objects in our galaxy, hosting processes far exceeding the energies reached in terrestrial, artificial accelerators. The exact nature and relationship of these objects with the galactic environment is still poorly understood; in particular the ``knee'' in the CR spectrum around $\sim 3 \:\rm{PeV}$ suggests a cutoff or transition in the astrophysical systems producing CRs. Direct investigation of the source of these CRs at extreme energies is complicated by redirection of charged CR particles by interstellar magnetic fields over galactic distance. Understanding these astrophysical system requires Very High Energy (VHE, $E\sim 0.1-100$~TeV) and Ultra High Energy (UHE, $E> 100$~TeV) gamma ray telescopes and neutrino detectors that detect uncharged particles that travel unimpeded from galactic sources to Earth.

The Large High Altitude Air Shower Observatory (LHAASO, \cite{LHAASObook}) observes VHE and UHE gamma rays with energies well into the PeV range. Over the past few years, LHAASO has detected 90 sources, 43 reaching UHE energies \citep{1LHAASO}. LHAASO has two component detector arrays; the Water Cherenkov Detector Array (WCDA) operating in $0.1 - 30 \: \rm{TeV}$, and the Square Kilometer Array (KM2A) in $30 \:\rm{TeV} - 30 \:\rm{PeV}$. 

In this work, we examine one particular region of VHE emission near $\rm{RA} = 55.5^\circ$, $\rm{dec} = 53.1^\circ$. Originally described in \cite{J0343paper} as a single extended source named LHAASO J0341+5258, further analysis with additional data and better sensitivity discerned two KM2A and one WCDA source in this region of the galactic plane \citep{1LHAASO}. In the first LHAASO catalog, 1LHAASO J0343+5254u (previously LHAASO J0341+5258) has a reported test statistic for photons above $100\:\rm{TeV}$ of $20.2$, suggesting that the underlying astrophysical system is a PeVatron (due to gamma ray emission above $100 \:\rm{TeV}$) and leading to substantial interest in followup observations and analyses.

In the original discovery paper \citep{J0343paper} and in subsequent, independent analyses \citep{J0343Fermi,J0343VERITAS}, special attention was given to a nearby \textit{Fermi}-LAT point source, 4FGL J0340.4+5302. In the 4FGL-DR4 catalog \citep{4FGLDR3,4FGLDR4}, 4FGL J0340.4+5302 is ``unassociated'', having no known counterpart at other wavelengths. Because of the sharp cutoff in the \textit{Fermi}-LAT spectrum around $\sim 2 \:\rm{GeV}$, these previous papers have treated this \textit{Fermi} source as a possible gamma ray pulsar, but without radio pulsations or gamma ray timing solution this classification is tentative. \cite{J0343Fermi} discussed the \textit{Fermi}-LAT gamma ray emission at 4FGL J0340.4+5302 and made a case for that pulsar-like source to be the lower-energy counterpart to the LHAASO source, establishing an upper limit on X-ray flux of approximately $10^{-12}~ \rm{erg/s/cm^2}$ at the \textit{Fermi} source using a brief archival Chandra data. Fortunately, there are two \textit{Swift}-XRT \citep{SwiftXRT} observations of this \textit{Fermi} source from 2012 totaling $\sim 5 \: \rm{ks}$ available in the HEASARC archive, which can give additional context to this \textit{Fermi}-LAT source.

Summing these two observations with \verb|XImage v4.5.1|, we find no X-ray point or extended source in the $95\%$ \textit{Fermi}-LAT uncertainty ellipse, for a $3\sigma$ upper limit for $0.3-10.0 \:\rm{keV}$ absorbed flux of $\approx 1 \times 10^{-13} \:\rm{erg/s/cm^2}$ (assuming $\Gamma_X = 2$). Notably, this upper limit is below the X-ray synchrotron flux predicted by \cite{J0343Fermi} in both their leptonic and hadronic models for the combined multiwavelength spectra. Ongoing work on additional Chandra observations that partially overlaps with the 95\% ellipse of 4FGL J0340.4+5302 has detected dim X-ray point sources with fluxes of $\sim 10^{-13} \:\rm{erg/s/cm^2}$ in the \textit{Fermi}-LAT region (Acero et al. in prep, private correspondence), which will certainly increase our understanding of 4FGL J0340.4+5302 as a component of the complex gamma ray landscape near 1LHAASO J0343+5254u.

It is still uncertain what electromagnetic and CR emission model is most appropriate for UHE sources like 1LHAASO J0343+5254u. Leptonic models -  dominated by electron synchrotron, inverse Compton (IC) upscattering of ambient photons by those electrons, or synchrotron self-Compton (SSC) - can explain a wide range of high-energy behaviors using a single population of parent particles in systems like PWN. Alternatively, hadronic processes like pion decay and CR interactions in supernova remnants can account for VHE and UHE emission, and hadronic processes are required to produce the neutrino emission that is suspected to accompany galactic PeVatrons. The discovery of a neutrino flux contained in the plane of the Milky Way by the IceCube observatory \citep{IceCubeMW} suggests that hadronic processes are at work in galactic VHE/UHE systems, as neutrinos are produced in copious amounts only by hadronic processes.

A notable challenge for understanding the multiwavelength behavior of galactic PeVatrons is the difficulty in confident association with lower-energy sources.  Not only are the uncertainty ellipses or spatial extension of UHE sources quite large, but their location in the galactic plane guarantees a bevy of gamma- and X-ray sources coincident but unrelated to the VHE and UHE emission. Furthermore, it is entirely possible for a source like a pulsar to be bright enough for detection in gamma rays but too dim for detection in X-rays or vice versa \citep{FermiPulsars}, so an X-ray source may not necessarily be related to a nearby VHE/UHE emission region. Special care should therefore be taken to coordinate low-energy followup to gamma ray emission to search for possible counterparts.

In this work, we report on XMM-Newton \citep{NewtonXMM} observations from Feb 2024 targeting 1LHAASO J0343+5254u (observation IDs 0923400401, 0923400801, and 0923401401). In section \ref{sec:ObsAna} we describe our observations, data reduction, and fitting of X-ray and gamma ray data. In section \ref{sec:MWLMod} we discuss multiwavelength leptonic modeling, and in section \ref{sec:Discussion} we discuss our X-ray and MWL modeling results and a related radio project. Finally, in section \ref{sec:Conc} we present our conclusions and discuss next steps.

\section{Observations and Data Reduction}
\label{sec:ObsAna}

\begin{figure*}[t]
    \centering
    \includegraphics[width=\textwidth]{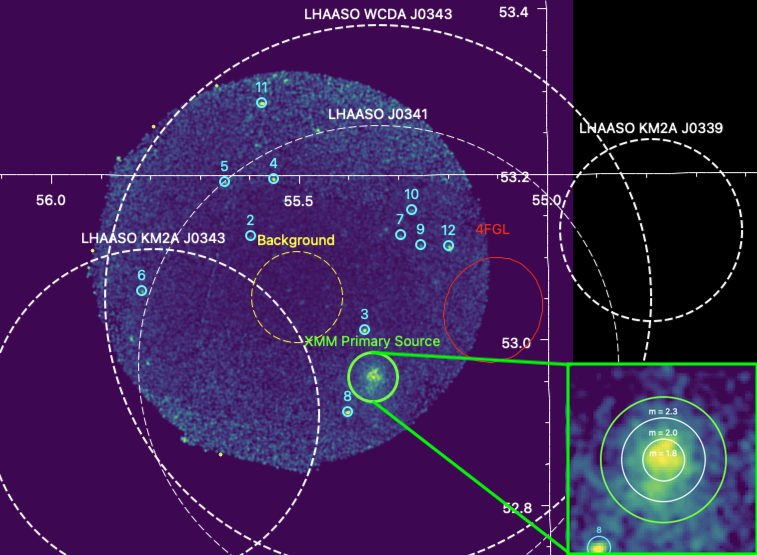}
    \caption{The summed XMM-EPIC MOS1 and MOS2 rate map, showing the LHAASO regions (white dashed - LHAASO J0341 is the source described in \cite{J0343orig}, the others are the KM2A and WCDA sources from \cite{1LHAASO}), the primary source region (green), the selected background region (yellow dashed), the nearby 4FGL 95\% uncertainty ellipse (red), and the eleven detected secondary point sources (cyan numbered). The lime insert shows the primary source in greater detail along with the fitted absorbed power law slopes in each annular analysis region.}
    \label{fig:XMMregions}
\end{figure*}

We downloaded the three observations (0923400401, 0923400801, and 0923401401) from the XMM-Newton Science Archive\footnote{\hyperlink{https://nxsa.esac.esa.int/nxsa-web/}{XMM-Newton Science Archive}, data becomes public on 25 March 2025}, and used the standard XMM-Newton analysis pipeline implemented in Python \verb|pysas v21.0.0| \citep{XMMSAS} to produce refined event files for each observation. Starting with \verb|epproc| and \verb|emproc|, we produced first-look light curves of the entire XMM-EPIC field of view to evaluate particle flaring times and establish a desirable rate cut.  We found that unfiltered per second event rates below $40 \:\rm{cps}$ for PN data and below $4 \:\rm{cps}$ for MOS1 and MOS2 data excluded particle flares and times with high background. A manual time interval filter was only necessary for observation 0923400801, due to a low plateau in event rate surrounding a smaller flare. These rate cuts create $87.7 \:\rm{ks}$ of good time between the three observations.

Applying these rate cuts, we used \verb|tabgtigen| to create good time intervals for each observation with additional filters, using event patterns 0 through 4 and energies $0.2 \:\rm{keV} < E < 15 \:\rm{keV}$ for PN data and patterns 0 through 12 and energies $0.2 \:\rm{keV} < E < 12 \:\rm{keV}$ for MOS1 and MOS2 data. Finally, we used \verb|evselect| to create cleaned, filtered event lists for each observation.

Examining the XMM-EPIC-MOS2 data by eye, we noticed a substantial extended region of X-ray emission in the south-west corner of the 1LHAASO J0343+5254u WCDA region. By XMM-\textit{Newton} conventions, this ``primary'' source is given the name XMMU 034124.2+525720, shown in the insert in Figure \ref{fig:XMMregions}.

In the XMM-Newton EPIC-MOS2 field of view, we noted several secondary X-ray point sources besides the extended primary region. These secondary sources deserve additional attention, especially as they are scattered across the entire region of interest for 1LHAASO J0343+5254u. Using the exposure-corrected and filtered EPIC-MOS2 data, we use the \verb|XIMAGE| routines \verb|detect| and \verb|sosta| to identify all sources with S/N$>4$ in the MOS2 data, finding eleven in the field of view upon excluding the extended primary source region described above. We label these sources 2 through 12 in no particular order.

To generate the spectrum of the primary source, we used a circular selection region centered on the maximum count rate in the primary region, at $\rm{RA} = 55.3509^{\circ}$, $\rm{Dec} = 52.9527^{\circ}$. We set the size of the primary region at $0.03^{\circ} = 1.8 \arcmin$, encompassing the entirety of the extended emission while not extending to either of the nearby point sources.  This size is similar to other X-ray PWN related to TeV emission regions like those linked to G75.2+0.1 \citep[][having X-ray size $\approx 1\arcmin$]{DragonflyPWN} and HAWC J1826-128 \citep[][$\approx 3 \arcmin$]{EelPWN}, an indication that this new source may be a PWN like those objects.

We select a nearby background region distant from the primary source in order to dodge any contamination from the outer limits of the primary extended emission. The background region was selected to be entirely contained in both the EPIC-MOS1 and -MOS2 fields of view, located at $\rm{RA} = 55.5050^{\circ}$, $\rm{Dec} = 52.0523^{\circ}$ with radius $0.055^{\circ}$. Using \verb|evselect|, we extracted background and source spectra for the primary and secondary sources using the clean, filtered event lists, generating \verb|.arf| and \verb|.rmf| files with \verb|arfgen| and \verb|rmfgen| respectively. 

We used \verb|epicspeccombine| to separately create summed EPIC-MOS and -pn spectra (it is not feasible to merge EPIC-MOS and EPIC-pn data into a single file). For some of the secondary point sources, their spatial location placed them outside of the XMM-EPIC-MOS1 field of view, so only -MOS2 data was used for those. Finally, we used the \verb|ftools| function \verb|ftgrouppha| to group each spectrum with minimum $S/N > 3$ in each grouped energy bin. This final step reduces the number of bins for fitting but guarantees that each energy bin individually would be a substantial detection and that each bin has approximately Gaussian errors appropriate for $\chi^2$ fitting.

\subsection{Primary Source Extension}
\label{sec:primext}

To better constrain the spatial extension of the primary source, we used the surface brightness analysis tool in \verb|ds9| to extract a surface brightness profile in the four cardinal directions from the center of the primary source in the summed EPIC-MOS count rate data. We extract the per square arcsecond count rate in rectangular tracks going north, east, south, and west from the centroid of the primary source. Each rectangular track has width $0.6 \arcmin$ to average out small-scale variations.

\begin{figure}
    \centering
    \includegraphics[width=\columnwidth]{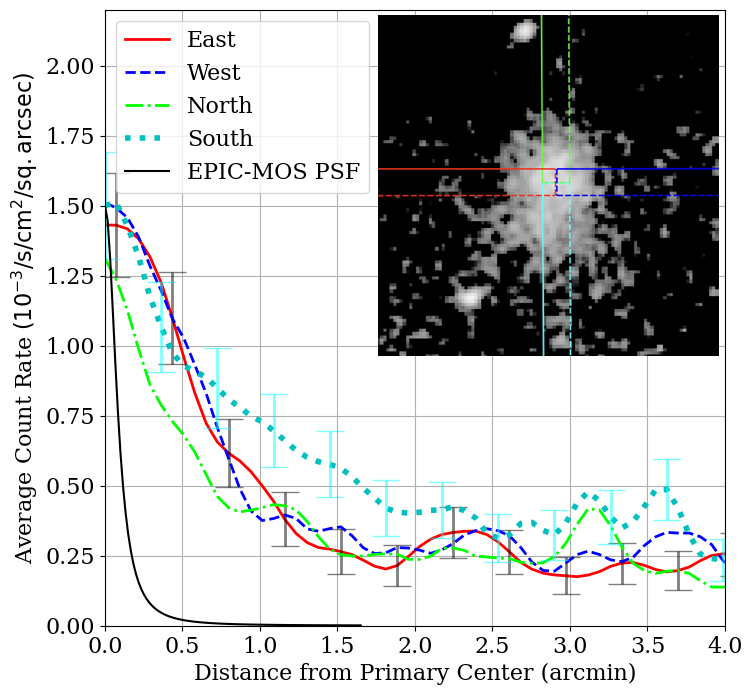}
    \caption{The average flux density along four tracks extended in the cardinal directions (East - red solid : West - blue dashed : North - lime dot-dashed : South - cyan dotted) from the primary source center, along with the EPIC-MOS PSF for a $10 \arcmin$ offset and $3 \:\rm{keV}$ energy (black). The insert shows the regions used in the extraction of these count rates. Uncertainties (black for east, cyan for south) assume Poisson statistics and are displayed every $0.36 \arcmin$.}
    \label{fig:NiceCountRate}
\end{figure}

Figure \ref{fig:NiceCountRate} shows that the surface brightness of the extended primary source drops down to background levels around $\sim 1 \arcmin$ from the centroid position. However, the southern portion of the primary source extends further, out to $\sim 2 \arcmin$, while the northern portion has a steeper drop in surface brightness than any other direction, establishing the asymmetric spatial shape of the primary source. The EPIC-MOS PSF is estimated using the XMM-Newton calibration file for the MOS camera (XRT1\_XPSF\_0016.CCF) using the King profile description at a $10 \arcmin$ distance from the center of the field of view at 3 keV. It is much smaller than the extension seen for the primary source as shown by the black line in Figure \ref{fig:NiceCountRate}. 

\subsection{Primary Source Fitting}
\label{sec:primfit}

Using \verb|Xspec v.12.14.0h| \citep{Xspec}, we conducted an absorbed power law fit to the background-subtracted primary source spectrum for each individual observation and for the summed spectra. For this fitting, we load both the EPIC-MOS and EPIC-pn data into \verb|xspec| and conduct a simultaneous fit of the -MOS and -pn data. There is an instrumental silicon feature appearing in the EPIC-MOS data as a deficit in the spectrum at approximately $1.7 \:\rm{keV}$, and an instrumental $8\:\rm{keV}$ feature in the EPIC-pn data. These features are due to the background spectrum of the cameras; by selecting a background region in a different CCD panel than the source, an instrumental excess feature in the background CCD becomes a deficit in the spectrum. We mask out data points between $1.65$ and $1.75 \:\rm{keV}$ for EPIC-MOS and between $7.5$ and $8.5 \:\rm{keV}$ for EPIC-pn to avoid any impact on our results.

The fit was of form \verb|tbabs * cflux * powerlaw|, with \verb|tbabs| giving X-ray absorption by the ISM, \verb|cflux| calculating the $0.2 - 12.0 \:\rm{keV}$ energy flux, and \verb|powerlaw| a simple power law with photon index $\Gamma_X$. For this fit, the galactic $n_H$ column density was started at the catalogued value ($n_H = 0.741 \times 10^{22} \:\rm{/cm^2}$ \cite{nhMap}) but allowed to vary as a free parameter. For \verb|tbabs|, we use cross-section data from \cite{Verner1996} and abundance calibration from \cite{Wilms2000}.

\begin{figure}
    \centering
    \includegraphics[width=\columnwidth]{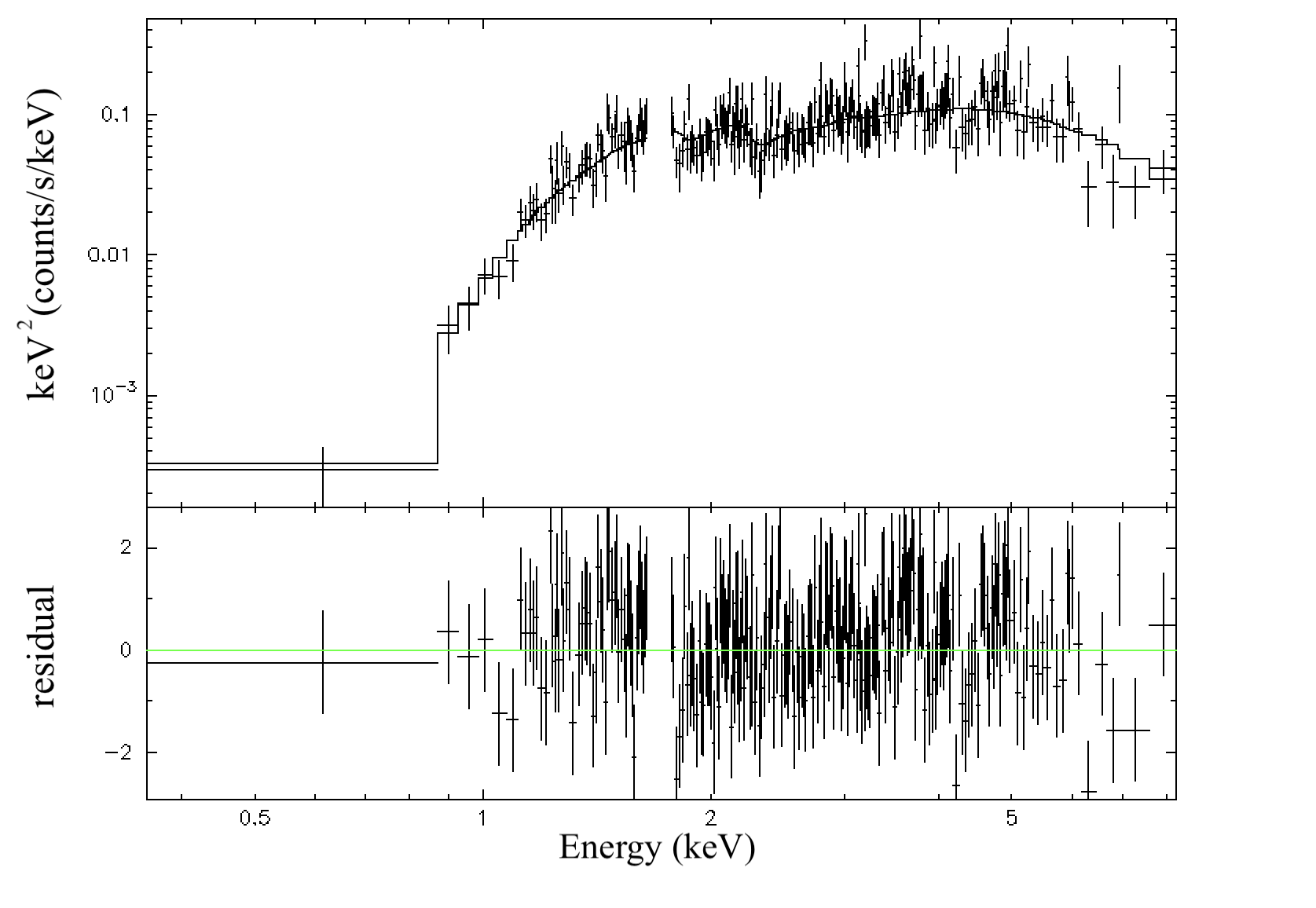}
    \caption{$0.3 - 8.0 \: \rm{keV}$ XMM-Newton spectrum of the primary source, fitted with an absorbed power law. An instrumental feature of the EPIC-MOS detector at $1.7 \:\rm{keV}$ is masked out for our fitting.}
    \label{fig:AllSpec}
\end{figure}

The spectrum of the primary source, shown with a power-law fit for the summed observations in Figure \ref{fig:AllSpec}, appears as a featureless power law with photon index $\Gamma_X = 1.9 \pm 0.1$. The fitted parameters for the primary source are shown in Table \ref{tab:fitPrim}, with $95\%$ uncertainties generated by the \verb|error| command. The fits from each individual observation are within errors of the summed fit, so there is no evidence to suggest any inter-observation variability in the X-ray spectrum of this source. We find no evidence for a high-energy cutoff or break in the power-law spectrum of the primary source, suggesting that the peak of the $\nu F_\nu$ flux is near or above the top of XMM's energy band.

\begin{deluxetable*}{lccccccc}
\tablecaption{Spectral fitting results for the primary source.} \label{tab:fitPrim}
\tablewidth{\columnwidth}
\tablehead{
\colhead{} & \colhead{Exposure} & \colhead{$\chi^2$} & \colhead{D.o.F.} & \colhead{$nh$} & \colhead{$\Gamma_X$} & \colhead{$F_X/10^{-12}$} \\
\colhead{} & \colhead{ks} & \colhead{} & \colhead{} & \colhead{$10^{22} cm^{-2}$} & \colhead{} & \colhead{$(\rm{erg/s/cm^2})$} }
\startdata
Initial & & & & $0.741$ & $2$ & 1 \\
All Observations & 87.7 & $497.76$ & $491$ & $1.68\pm0.12$ & $1.91\pm0.08$ & $ 1.29 \pm 0.05 $ \\
0923401401 & 43.0 & $601.45$ & $572$ & $1.65\pm0.18$ & $1.90\pm0.12$ & $ 1.24 \pm 0.06 $ \\
0923400401 & 21.6 & $287.96$ & $304$ & $1.75\pm0.26$ & $1.98\pm0.18$ & $ 1.25 \pm 0.09 $ \\
0923400801 & 23.1 & $386.61$ & $342$ & $1.81\pm0.31$ & $1.98\pm0.20$ & $ 1.18 \pm 0.10 $ \\
\enddata
\end{deluxetable*}


To test whether $\Gamma_X$ varies based on distance from the central, brightest part of the primary source, we also subdivided the primary source into annular regions concentric with the center of the region and performed spatially resolved spectral analysis.  We created a core region of radius $0.6 \arcmin$, an inner annulus beginning there and extending to $1.2 \arcmin$, and an outer annulus from $1.2 \arcmin$ to $1.8 \arcmin$, shown in the insert of Figure \ref{fig:XMMregions}. We conducted identical power law fits using all XMM-EPIC data for these three separate regions. We fix $n_H=1.68 \times 10^{22} \:\rm{/cm^2}$ the value from the overall fit.

For the core, inner, and outer regions, we find photon indices $\Gamma_X$ of $1.80\pm0.09$, $1.99\pm0.09$, and $2.27\pm0.15$, respectively. These fits show a statistically significant difference between the power law slopes of the core part of the primary source and the annular outer part, demonstrating that the outermost regions of the primary source have notably softer X-ray spectra. 



To facilitate multiwavelength fitting of this system, we extract de-absorbed $E^2 \frac{dN}{dE}$ fluxes for the fully fitted all-observation primary model by setting $n_H = 0$ and using four logarithmic energy bins between $2 \:\rm{keV}$ and $8 \:\rm{keV}$, with proportional uncertainties obtained via \verb|xspec| for the model flux. These fluxes can be incorporated into a wider SED spectrum and compared to other fluxes directly.

Conducting timing analysis of the innermost $0.6 \arcmin$ region of the primary source, we detect no evidence for X-ray pulsations coming from the region. However, the PN-mode data taken by XMM-Newton is not in the mode most appropriate for superior timing sensitivity, so additional X-ray observations with more discerning timing registration are warranted.

\subsection{Secondary Sources}

We fit the EPIC-MOS data for each secondary source with an absorbed power law model to evaluate the general shape of each spectrum. In this case we kept $\rm{n_H}$ fixed to the previous fitted value of $1.68 \times 10^{22} \:\rm{/cm^2}$. Cataloged $n_H$ values do not substantially vary over the field of our observations, though some of these secondary X-ray sources may be extragalactic, which could impose additional $n_H$ column density. For a first-look flux, we fixed $\Gamma_X = 2$, and subsequently freed that parameter for a full fit. We also note the $0.2-12.0 \:\rm{keV}$ background-subtracted count rate for each source, which is a more observational measurement of the photon flux from each secondary source.

In Table \ref{tab:secondary}, we report the fitted fluxes and power-law slopes for each of the detected secondary sources.  Some of the dimmer sources have very soft fitted spectral indices, leading to unrealistic and disproportionate values for their flux; with such a steep spectrum, more photons would be emitted below $1 \:\rm{keV}$ where X-ray absorption is substantial, so the calculated unabsorbed flux via \verb|cflux| would be unrealistically high. The reported count rate in Table \ref{tab:secondary} is a reasonable proxy for flux in these cases.

Conducting a cross-reference with the 2 Micron All-Sky Survey \citep{2MASS} and with the SIMBAD database, none of the secondary sources are coincident with IR source down to $m_J = 15$ or SIMBAD optical sources, with the exception of source 8, south-east of the primary. That source in particular is coincident with the star UCAC4 715-027962 of magnitude $V=12.5$, listed as a spectroscopic binary in SIMBAD. This star is near the brightness that would incur optical loading in the XMM-EPIC detector in full-frame mode, so its detection as an X-ray source is suspect. Additionally, using a general cross-reference with the SIMBAD catalog, source 7 is noted as a ROSAT X-ray point source. Its reported flux in the ROSAT dim source catalog \citep{ROSATdim} is substantially higher than our observed flux in XMM, suggesting a degree of X-ray variability. These secondary sources may be interesting targets for further follow-up.

\begin{deluxetable*}{lccccccl}
\tablecaption{Observational parameters, background-subtracted count rate in XMM-EPIC, and power law fits for the secondary sources.} \label{tab:secondary}
\tablewidth{\columnwidth}
\tablehead{
\colhead{ID} & \colhead{RA (J2000)} & \colhead{Dec (J2000)} & \colhead{Count rate} & \colhead{$\log F_X$, $\Gamma_X = 2$}& \colhead{$\log F_X$} & \colhead{$\Gamma_X$} & \colhead{Notes} \\
\colhead{} & \colhead{HH:MM:SS} & \colhead{dd:mm:ss} & \colhead{$1/\rm{ks}$} & \colhead{$\rm{erg/s/cm^2}$} & \colhead{$\rm{erg/s/cm^2}$} & \colhead{} & \colhead{} }
\startdata
2 & 03:42:23.2 & 53:07:34.9 & $1.2 \pm 0.2$ & $-13.5 \pm 0.1 $ & $-13.5 \pm 0.1 $ & $1.8 \pm 0.5 $& \\
3 & 03:41:28.3 & 53:00:45.4 & $4.2 \pm 0.2$ & $-12.9 \pm 0.1 $ & $-12.9 \pm 0.1 $ & $1.7 \pm 0.2 $ & \\
4* & 03:42:12.1 & 53:11:42.4 & $1.0 \pm 0.1$ & $-13.1 \pm 0.1 $ & $-13.2 \pm 0.1 $ & $1.4 \pm 0.7 $ & EPIC-MOS2 only\\
5 & 03:42:35.5 & 53:11:30.2 & $1.5 \pm 0.2$ & $-13.2 \pm 0.1 $ & $-13.1 \pm 0.2 $ & $1.0 \pm 0.7 $ & \\
6 & 03:43:15.7 & 53:03:33.4 & $0.8 \pm 0.2$ & $-12.8 \pm 0.2 $ & $-13.0 \pm 0.4 $ & $1.2 \pm 2.0 $ & \\
7 & 03:41:10.8 & 53:07:35.1 & $1.0 \pm 0.2$ & $-13.4 \pm 0.2 $ & $-13.3 \pm 0.5 $ & $0.3 \pm 1.4 $ & ROSAT X-ray Source\\
8 & 03:41:36.5 & 52:54:49.8 & $3.4 \pm 0.2$ & $-13.0 \pm 0.1 $ & $-12.0 \pm 0.3 $ & $4.0 \pm 0.5 $ & $V=12.5$ optical loading\\
9 & 03:41:01.0 & 53:06:54.3 & $0.7 \pm 0.2$ & $-13.3 \pm 0.2 $ & $-12.5 \pm 1.5 $ & $3.21 \pm 1.9 $ & \\
10 & 03:41:05.1 & 53:09:26.0 & $0.8 \pm 0.2$ & $-13.4 \pm 0.2 $ & $-13.0 \pm 0.6 $ & $0.2 \pm 1.6 $ & \\
11* & 03:42:17.8 & 53:17:12.6 & $0.6 \pm 0.1$ & $-13.0 \pm 0.2 $ & $-13.1 \pm 0.4 $ & $0.9 \pm 0.9 $ & EPIC-MOS2 only\\
12 & 03:40:47.4 & 53:06:48.7 & $2.1 \pm 0.2$ & $-13.1 \pm 0.1 $ & $-13.1 \pm 0.1 $ & $1.9 \pm 0.4 $ & \\
\enddata
\end{deluxetable*}

\subsection{Fermi-LAT Data Analysis}
\label{sec:Fermi}

We also analyzed \textit{Fermi} Large Area Telescope (\textit{Fermi}-LAT) photon data in the sky region surrounding the XMM-Newton position. The nearby \textit{Fermi} 4FGL source J0340.4+5302 is not spatially coincident with our XMM-Newton primary source, but could be related to a relic radio PWN related to our candidate X-ray PWN by the movement of a pulsar with a natal kick. We evaluate the spatial extent of the 4FGL source and obtain upper limits at the XMM-Newton position. We performed data reduction, event selection, and time filtering using \texttt{fermitools} version 2.2.0 and \texttt{fermipy} version 1.1.6 \citep{FermiPy}.

We used \textit{Fermi}-LAT photon data from August 2008 to August 2024 in the $0.5-500 \:\rm{GeV}$ energy range in a circular region with radius $20^{\circ}$ centered at the XMM-Newton primary source. We performed a summed likelihood analysis using the PSF event types, splitting Fermi-LAT events according to their quality of direction reconstruction (PSF0 being the worst reconstruction). We used the following selection as a function of energy: PSF2+3 from 0.5-1 GeV and PSF0+1+2+3 for 1-500 GeV, to keep only the best reconstructed events at low energies. The Fermi-LAT instrument response function version P8R3-SOURCE-V3 was used in the analysis, with maximum value of the zenith angle of observation was 90$^{\circ}$. 

We included all 4FGL-DR4 \citep{4FGLDR4} gamma ray point sources and extended sources in the gamma ray background model. For the diffuse background model, we used the Galactic diffuse emission template (\texttt{gll\_iem\_v07.fits}) and the relevant extra-Galactic isotropic emission templates for each PSF type (in the form of iso\_P8R3\_SOURCE\_V3\_PSF*\_v1.txt, where * is the PSF type described above) obtained at the Fermi science center \citep{LATmodels}.

\begin{figure}
    \centering \vspace*{1pt}
    \includegraphics[width=\columnwidth]{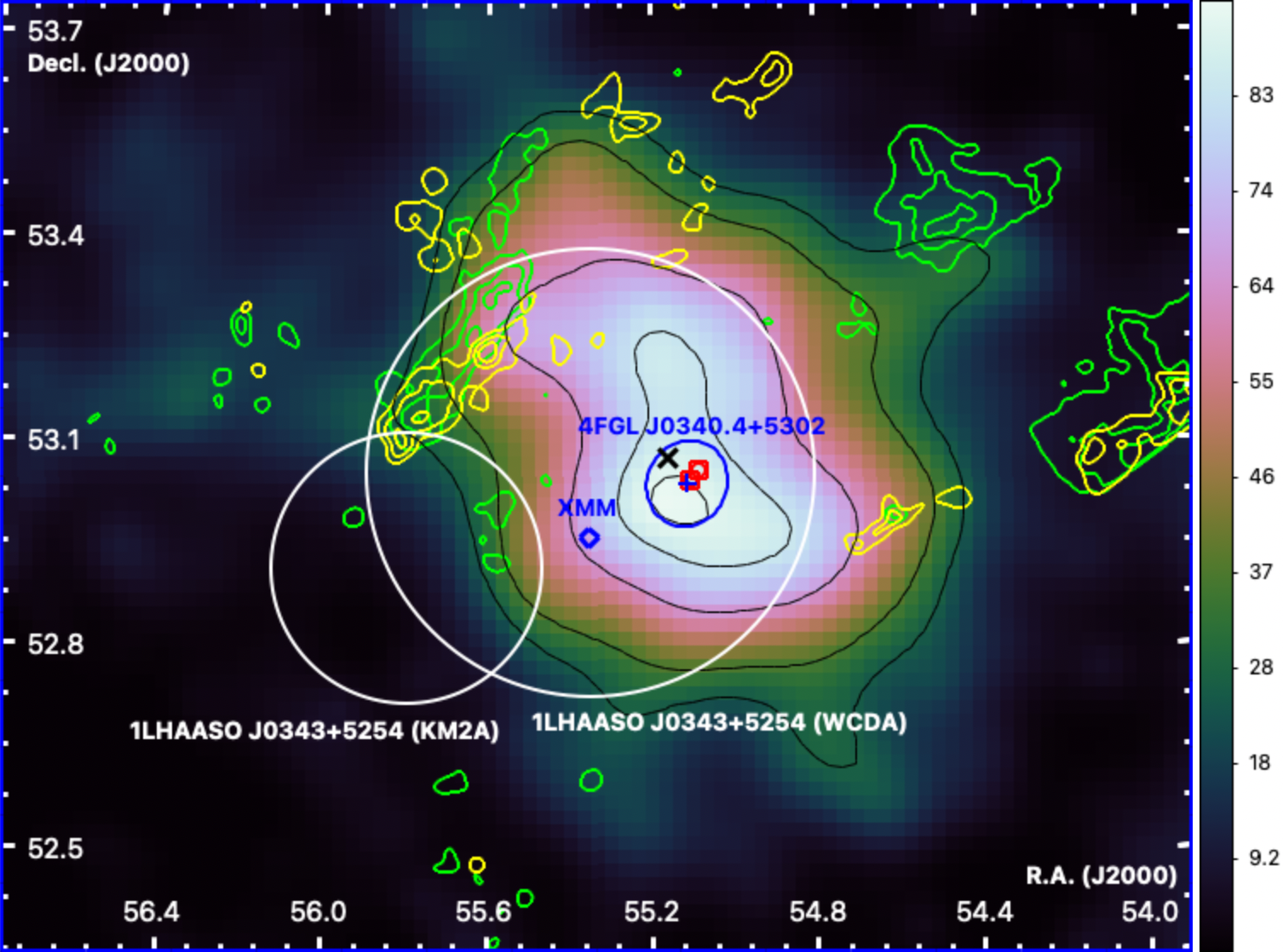}
    \caption{\footnotesize{$0.5 - 500 \:\rm{GeV}$ \textit{Fermi}-LAT TS map after removing 4FGL J0340.4+5302 from the gamma ray background model. We include the 1LHAASO J0343+5254u WCDA and KM2A regions (white), the 95\% error ellipse of 4FGL J0340.4+5302 (blue ellipse), and the extent of the XMM-Newton primary source (small blue circle). The red point are two X-ray point sources described in upcoming works. Nobeyama radio CO line contours are shown for velocity range of [-20, -4] (yellow) and [-4, +10] (green) km/s.
     }}
    \label{fig:TSMaps_fullE}
\end{figure}

To visualize the spatial morphology of 4FGL J0340.4+5302, we deleted that source from the gamma ray background and reproduced the test statistic (TS) map in Figure \ref{fig:TSMaps_fullE}.  Besides the 95\% uncertainty ellipse of 4FGL J0340.4+5302, Figure \ref{fig:TSMaps_fullE} also shows additional excess gamma ray emission around the positional error ellipse of 1LHAASO J0343+5254u in white.



We measured the extension of 4FGL J0340.4+5302 with the \texttt{extension} tool under two different spatial models, \verb|RadialGaussian| and \verb|RadialDisk|, measuring a likelihood ratio of extension $\rm{TS}_{\rm{ext}} = -2 \log{L_{\rm{ext}}/L_{\rm{pt}}}$. If $\rm{TS}_{\rm{ext}} > 16$, the tested source is likely extended.  The models have $\rm{TS}_{\rm{ext}} = 2.19$ and $= 2.18$ respectively, suggesting that 4FGL J0340.4+5302 is a point source in the energy range of 500 MeV - 500 GeV. We did find that $\rm{TS}_{\rm{ext}}$ increased substantially depending on the low-energy limit of our event selection, suggesting that there may be substantial energy dependence on the morphology of gamma ray emission in this region.



The best-fit position of the point-like source with a LP-type spectrum was calculated using the \texttt{localize} method from the \texttt{fermipy} package. The best-fit position of 4FGL J0340.4+5302 for 500 MeV - 500 GeV was found to be R.A.(J2000), Decl.(J2000) = 55$^{\circ}\!$.1426 $\pm$ 0$^{\circ}\!$.0174, 53$^{\circ}\!$.0656 $\pm$ 0$^{\circ}\!$.0181, which is compatible with the location reported for 4FGL J0340.4+5302 in the 4FGL-DR4 catalog. 


In order to model the gamma ray flux at the position of our XMM-Newton primary source, we use a two-source model for \textit{Fermi}-LAT emission in the region, with one source at the position of 4FGL J0340.4+5302 and another at the position of our XMM-Newton primary source. The gamma ray spectrum of 4FGL J0340.4+5302 was modeled as a log-parabola (LP), the model best fitting its spectrum in the 4FGL-DR4 catalog \cite{4FGLDR4}, with the form


$$ \frac{dN}{dE} = N_0 \left(\frac{E}{E_0} \right)^{-(\alpha + \beta \log{(E/E_b)})}$$

\noindent with ${\rm E_0}$ the reference energy and $E_b$ the break energy. The spectral indices are ${\rm \alpha}$ and ${\rm \beta}$. The hypothetical additional source at the XMM-Newton position was modeled with a power-law with $\Gamma_\gamma = 2$



In the two-point-sources scenario, 4FGL J0340.4+5302 was detected with a TS value of $\sim 834$ ($28.9\sigma$), with spectral parameters $\alpha = 3.51 \pm 0.27 $  and $\beta = 0.747 \pm 0.352 $. The total energy flux was found to be $ (5.4 \pm 0.3) \times 10^{-6} \:\rm{MeV cm^{-2} s^{-1} } $. No significant gamma ray emission was detected at the position of the XMM-Newton primary source, so we evaluated upper limits for that position by assuming a power-law spectrum with $\Gamma_\gamma = 2.0$. We find an upper limit on $0.5-500 \:\rm{GeV}$ flux at the position of the XMM-Newton primary source of $6.9 \times 10^{-10} \:\rm{erg/s/cm^2}$, and also establish upper limits in smaller energy ranges, shown in Figure \ref{fig:nicespec}.



\section{Multiwavelength Modeling}
\label{sec:MWLMod}

\subsection{MWL Spectrum Construction}

The primary XMM-Newton source has a featureless power law X-ray spectrum for the overall region, and asymmetrical spatial extensions reaching arcminutes away from a central bright spot. These X-ray properties point towards classification as an X-ray PWN, similar to other systems like HAWC J1826-128 (the Eel, \cite{EelPWN}), G106.3+2.7 (the Boomerang, \cite{BoomerangPWN}), and G75.2+0.1 (the Dragonfly, \cite{DragonflyPWN}). Given the complex gamma ray sky around 1LHAASO J0343+5254u, it is possible that the XMM primary source is related to the high-energy emission observed by LHAASO like the other gamma ray PWN noted above.

To create a MWL spectrum and test whether the primary source is feasibly related to the LHAASO emission as a PWN, we join our measured X-ray flux points from Section \ref{sec:primfit} with \textit{Swift}-BAT upper limits from nondetection in \textit{Swift}-BAT \citep{SwiftBAT}, the derived \textit{Fermi}-LAT upper limits at the position of the primary source from Section \ref{sec:Fermi}, and the LHAASO spectrum from \cite{J0343orig}, to form an SED spanning nine orders of magnitude in energy. We also incorporate the VERITAS upper limits from \cite{J0343VERITAS} in our model, but those VERITAS upper limits are modeled in a smaller area of the sky containing only about $\sim 25 \%$ of the LHAASO source region. To be as conservative as possible while still including as much information as possible in our spectrum, we multiply the limits reported in \cite{J0343VERITAS} by four, which is a workaround that in all ways treats the limits as conservatively and cautiously as possible; the flux observable with VERITAS is certainly below our recalculated limits. These VERITAS limits, given these caveats, do not in the end affect the leptonic modeling described below.

The LHAASO spectrum reported in \cite{J0343orig} has been slightly updated in the first LHAASO catalog \citep{1LHAASO} with a flux and slope, but not with detailed points as in \cite{J0343orig}. The fluxes and slopes reported in \cite{1LHAASO} are similar to the overall spectrum described in \cite{J0343orig}, a work that included substantially more detailed explanations of the spectrum of this region. We adopt the spectrum from \cite{J0343orig}, acknowledging that the gamma ray emission has been decomposed into sub-sources in \citep{1LHAASO}.

\begin{figure*}
    \centering
    \includegraphics[width=\linewidth]{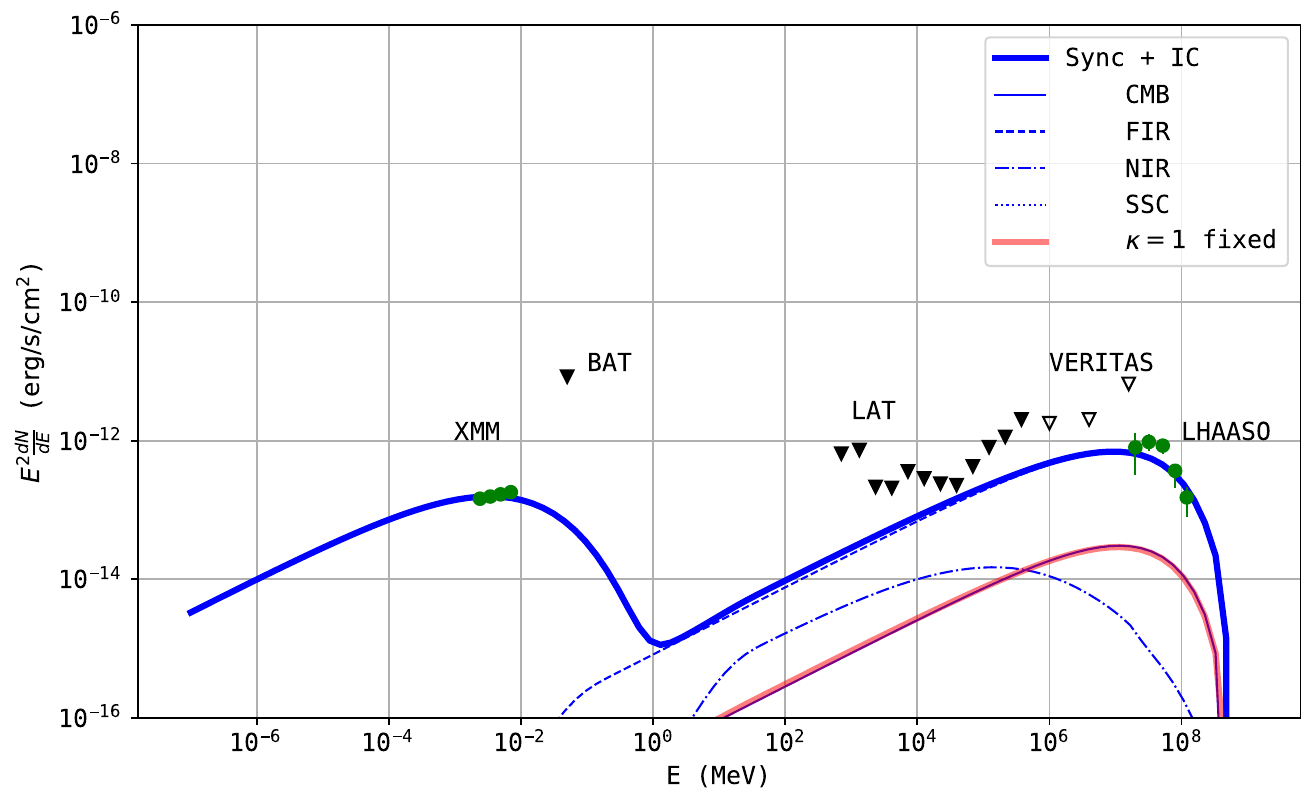}
    \caption{Detections (green dots) and constraining upper limits (black arrows) for our PWN candidate, along with the MCMC best-fit leptonic model (blue). The faint red curve is the leptonic IC spectrum, but with $\kappa = 1$, photon fields fixed at the solar neighborhood values. Components of the gamma ray emission are included as thin lines of the same color for each model (synchrotron self-Compton is not visible on this scale).}
    \label{fig:nicespec}
\end{figure*}

We manually searched optical and IR catalogs for low-energy point or extended counterparts to the candidate PWN but found no counterpart. The MWL spectrum constructed from the XMM-Newton and LHAASO detections and other upper limits is shown in Figure \ref{fig:nicespec}.

\subsection{Leptonic Model Fitting}

The electromagnetic emissions from systems like HESS J1825-137 \citep{HESSpwn}, the Eel \citep{EelPWN}, and the Boomerang \citep{BoomerangPWN} have been successfully modeled as fully leptonic. These models invoke electron synchrotron radiation in a magnetic field of magnitude $\sim 1 - 10 \:\rm{\mu G}$ for the X-ray radiation, plus gamma ray emission via inverse Compton scattering of ambient cosmic microwave background, near IR, and far IR photon fields (plus, to a lesser extent, synchrotron-self Compton upscattering). Conducting phenomenological fitting using a synchrotron+IC model can test whether there are feasible leptonic scenarios to explain the MWL emission from our candidate PWN.

We use the \verb|naima| python package \citep{Naima} to construct a leptonic model for the MWL spectrum and constraining upper limits. We adopt a exponential cutoff power law for the electron population, 

$$\frac{dN}{dE} = N_0 (E/E_0)^{-\alpha} e^{- \left( E/E_{\rm{cut,e}} \right)^\beta }$$ 

\noindent between electron energies of $100 \:\rm{MeV}$ and $10 \:\rm{PeV}$. $N_0$ is an amplitude for the electron distribution, and we set reference energy $E_0 = 1 \:\rm{TeV}$ and $\beta = 1$. The electron distribution has three free parameters: power law slope $\alpha$ and cutoff energy $E_{\rm{cut,e}}$, plus $N_0$. We fix the distance to the system to $1 \:\rm{kpc}$, a galactic distance similar to the characteristic distances of the spiral arms in the outer galaxy discussed in Section \ref{sec:GalEnv} and four of the newly discovered radio sources discussed in Section \ref{sec:Radio}.

The synchrotron emission occurs in a region with ambient magnetic field $B$ of angular size $0.5 \:\rm{pc} \frac{d}{\rm{kpc}}$ (the physical size corresponding to $2 \arcmin$ at distance $d$). The inverse Compton scattering \citep{ICFieldNaima} occurs in a wider populated with ambient cosmic microwave background and locally produced synchrotron X-ray photons (which only minimally contribute to upscattering), plus far IR (FIR, $20 \:\rm{K}$, $\rho_E = 0.3\kappa \:\rm{eV/cm^3}$) and near IR/optical (NIR, $5000 \:\rm{K}$, $\rho_E = 1.0\kappa\:\rm{eV/cm^3}$) photon fields. \verb|naima| integrates the upscattering of these photons by electrons with energy between $1 \:\rm{GeV}$ and $510 \:\rm{TeV}$. Initially, we fixed the intensity of these ambient photon fields to the solar neighborhood values. In that case, the red line in Figure \ref{fig:nicespec} shows that only $\sim 10\%$ of the observed LHAASO flux can be explained by IC emission in this case. Many other leptonic models allow the field intensity to be a free parameter, so we multiply the energy densities $\rho_E$ of the NIR and FIR fields by a final fitted parameter $\kappa$. For this IC regime the FIR field is substantially more relevant than the NIR field, and scaling both densities together ensures that the relative temperature of the overall field stays constant.

As a starting point for fitting, we adopt $\alpha = 2$, $E_{\rm{cut,e}} = 100 \:\rm{TeV}$, $B = 4.5 \:\rm{\mu G}$, and $\kappa = 50$. While this value for $\kappa$ means a substantially higher FIR field compared to the solar neighborhood, elevated FIR fields up to two orders of magnitude higher than local values are not unusual for leptonic models for IC emission from PWN \citep{EelPWN}, and can be achieved in star-forming or gas/dust rich regions of the galactic spiral arms.

To more fully explore the parameter-space of this model, we use the \verb|emcee| package to conduct a Markov Chain Monte Carlo (MCMC) ensemble sampling. This simulation uses ``walkers'', each starting at a slightly different initial condition, that iteratively explore the multidimensional parameter-space of the model. Eventually, the walkers produce an ensemble of parameter vectors which can constrain the best parameters for the model, visualized with a pairs plot. We use an MCMC ensemble sampling routine with 100 individual walkers, running for 200 steps after a 50 step burn-in, to characterize the parameter space of the leptonic model. 

For the leptonic fit, the MCMC sampler obtained a best fit of $\alpha = 2.0 \pm 0.2$, $E_{\rm{cut,e}} = (130 \pm 19) \:\rm{TeV}$, $B = (5.1 \pm 2.3) \:\rm{\mu G}$, and $\kappa = 61 \pm 15$ (amplification of the ambient IR fields by a factor of $61$ over the local value in the solar neighborhood). The uncertainty on $B$ is high due to degeneracies between $B$ and other parameters including $\alpha$ and $\kappa$. The amplitude of the lepton distribution is $N_0 = 7.0 \times 10^{30} \:\rm{/eV}$, for a total electron energy of $2.1 \times 10^{44} \:\rm{erg}$.  These parameters are summarized in Table \ref{tab:MCMCparam}, and this model is shown in blue in Figure \ref{fig:nicespec}.

\begin{deluxetable*}{lc|l}
\tablecaption{MCMC best fit parameters and $2\sigma$ uncertainties for the leptonic model} \label{tab:MCMCparam}
\tablewidth{\columnwidth}
\tablehead{ \colhead{Parameter} & \colhead{units} & \colhead{best fit} }
\startdata
$E_{tot,e}$ & $\rm{erg}$ & $(2.1 \pm 0.3) \times 10^{44}$ \\
$E_{\rm{cut},e}$ & $\rm{TeV}$ & $130 \pm 19$ \\
$\alpha$ & & $2.0 \pm 0.2$ \\
 & & \\
$B$ & $\rm{\mu G}$ & $5.1 \pm 2.3$ \\
$\kappa$ & & $61 \pm 15$ \\
\enddata
\end{deluxetable*}

\section{Discussion}
\label{sec:Discussion}

In terms of X-rays, the primary XMM source has spatial and spectral features much like other PWN like the Eel, Boomerang, and Dragonfly, all of which been linked to nearby $>\rm{TeV}$ emission regions observed in VERITAS \citep{BoomerangPWN}, HAWC, and HESS \citep{EelPWN}. In these systems, the nearby $>\rm{TeV}$ emission regions are generally more extended than the X-ray emission, which is not always centrally located in the gamma ray region, similar to our candidate PWN. In this way, our primary source may be linked to the wider LHAASO emission region by the same mechanisms.

A useful comparison is the extended pulsar wind nebula complex HESS J1825-137, described in detail in \cite{HESSpwn}. In that work, GeV and TeV emission extending out to $\sim 1^\circ$ around a region of X-ray emission ascribed to a PWN.  The X-ray emitting nebular region in that work has a compact core and extends asymmetrically to a few arcminutes, similar to our primary X-ray source.  In that system, the X-ray emission is ascribed to synchrotron radiation from $\sim TeV$ leptons in a $3 \mu G$ magnetic field, while gamma ray emission is produced from inverse Compton scattering of ambient photons off $> \rm{GeV}$ leptons further from the central X-ray region. A similar leptonic model could apply to the joint emission of our candidate PWN and the VHE/UHE emission observed with LHAASO. Though the emission from our candidate PWN does have some substructure shown in the insert in Figure \ref{fig:XMMregions}, our XMM observations do not have sufficient spatial resolution to decisively identify a point source in the candidate PWN that may be the pulsar itself.

In the 2021 paper that originally described gamma ray emission in this region \citep{J0343orig}, the LHAASO collaboration presented leptonic and hadronic fits to the LHAASO data, using only weak X-ray constraints and \textit{Fermi}-LAT data relating to the nearby 4FGL J0340.4+5302. To obtain an appropriate fit to the observed gamma ray cutoff, their used an exponential power law particle distribution with $\beta = 2$ as opposed to $\beta = 1$ in this work. Given that the 2021 LHAASO data will likely be updated, we do not comment on the wisdom of modifying $\beta = 1$, only noting that an updated LHAASO spectrum will help constrain whether a sharper cutoff in the particle spectrum is warranted. Allowing $\beta > 1$ would better fit the gamma ray cutoff presented in \cite{J0343orig} but would necessitate an increased $B$ field magnitude for the synchrotron emission as well. 

Extrapolating the models presented in \cite{J0343orig} to $\rm{keV}$ energies results in a predicted X-ray spectrum with a peak around $\sim 1 \:\rm{keV}$. Because our candidate PWN has an X-ray spectrum that must peak substantially higher, modifications to the models proposed in \cite{J0343orig} are warranted.

\subsection{Leptonic Model Analysis}

The blue line in Figure \ref{fig:nicespec} show the MCMC best-fit leptonic model, along with gamma ray constituent components in thinner lines of the same color. The faint red gamma ray curve shows the IC emission in the leptonic case if $\kappa = 1$, the NIR/FIR fields fixed at the solar value.  In the best-fit leptonic model, the values for $\alpha$, $E_{\rm{cut,e}}$, and $B$ are typical of PWN related to VHE emission regions like the Eel. As mentioned previously, an elevated $\kappa$ would be readily available in a dusty or gas-filled region of space like a star-forming region as optical light is reprocessed to lower energies. Low-energy observations near this new candidate PWN will more greatly constrain astronomical features that might contribute to the ambient IR field.



In the leptonic model, the limits prescribed by \textit{Fermi}-LAT mean that the electron distribution must have $\alpha \lesssim 2$, a value similar to other VHE PWN. The featureless X-ray spectrum observed in our XMM-Newton data constrains the magnetic field $B$ producing the synchrotron emission to be high, with $B \approx 5\:\rm{\mu G}$, akin to other VHE PWN, though there are substantial uncertainties in our fitted $B$ due to degeneracies in the model with other parameters. The magnetic field in the leptonic PWN model can be independently constrained by considering the size of the candidate PWN and the cooling time of the leptons diffusing throughout its volume. Equation 6 in \cite{Reynolds_2018} gives the synchrotron loss timescale $t_{1/2}$, the time for an electron primarily emitting at energy $E$ to lose half its energy in a magnetic field $B$.

$$ t_{1/2} = 1.2 \:\rm{kyr} \left( \frac{E}{\rm{keV}} \right)^{-1/2} \left( \frac{B}{10 \:\rm{\mu G}} \right)^{-3/2} $$

Though the great uncertainty in the distance to the candidate PWN prevents accurate measurement of its size, we can parameterize the diffusion timescale of leptons diffusing at velocity $v_{ad} \approx 1000 \:\rm{km/s} \approx 1\:\rm{pc/kyr}$ \citep{Porth_2016,Reynolds_2018}, using the angular size of the candidate PWN $\theta \approx 2 \arcmin$ and scaling for some distance $d$.

$$ t_{\rm{dif}} = \frac{\theta d}{v_{ad}} = 0.5 \:\rm{kyr} \left( \frac{d}{\rm{kpc}}\right)$$

Setting these two timescales equal to each other and solving for $B$ gives

$$ B = 18 \:\rm{\mu G} \left(\frac{d}{\rm{kpc}} \right)^{-2/3} \left(\frac{E}{\rm{keV}} \right)^{-1/3} $$

This equation presents an upper limit; if $B$ were to exceed the predicted value, leptons would not disperse through the observed angular size of the candidate PWN while cooling. For $d = 1 \:\rm{kpc}$ and $E = 10 \:\rm{keV}$ (the XMM spectrum still rising at $8 \:\rm{keV}$), we obtain an upper limit on $B$ of $B < 8.5 \:\rm{\mu G}$. This qualitative examination would suggests that the leptonic PWN model described above does not require unreasonable magnetic fields for a PWN of this size.

With $\kappa = 1$, no leptonic fit for the entire composite could be obtained, with the gamma ray emissivity being simply too low to account for more than $\approx 10\%$ of the observed LHAASO emission. Boosting $\kappa$ even moderately results in IC from the electrons of the PWN being a substantial if not dominant contributor to the observed LHAASO flux. It is also possible that some of the gamma ray emissivity observed with LHAASO, which is of interest to neutrino observatories like IceCube as a possible contributor to the galactic neutrino flux \citep{IceCubeMW}; neutrino emission is only expected from sources with a hadronic emission component.



In the leptonic model, the total energy in the primary particles is substantially lower than those energies reported for other systems \citep{EelPWN,DragonflyPWN}, suggesting that this candidate PWN is a substantially less luminous system than these other objects. In the leptonic model, $\log{E_{\rm{cut}}/\rm{TeV}} \approx 2$ and $1\:\rm{\mu G} < B < 10\:\rm{\mu G}$, values typical for PWN that are linked to VHE emission like the Eel, Dragonfly, or Boomerang. 

We only require a single astronomical accommodation to explain most if not all of the observed LHAASO gamma ray flux (elevated $\kappa$ to increase the ambient photon fields). The candidate PWN could contribute only part of the observed LHAASO gamma ray emission via IC upscattering, with additional leptonic or hadronic contributions making up the difference. At the moment we lack the detailed astronomical observations at other wavelengths needed to constrain the many free parameters in a leptonic+hadronic model; further radio, IR, and optical observations will provide further context for our candidate PWN and the broader environment of 1LHAASO J0343+5254u.

\subsection{Hadronic Models}

Though a leptonic model obtains an appropriate fit for this system and others like it, it is also possible that some of the gamma ray emissivity observed with LHAASO is hadronic in origin, via the decay of neutral pions produced by energetic protons interacting with ambient hadrons. Hadronic emission is of interest to neutrino observatories like IceCube as a possible contributor to the galactic neutrino flux \citep{IceCubeMW}; neutrino emission is only expected from sources with a hadronic emission component.

There are several astronomical unknowns that inhibit a detailed and tightly constrained hadronic model. First, lacking a evolutionary model for a pulsar in the heart of the candidate PWN prevents physical conclusions about the flow of particles into a target molecular cloud, the first step in a hadronic model. Speaking of, there is substantial uncertainty about the molecular cloud that would serve as a target; recent radio observations (discussed below) have only revealed several tentative cloud regions, with substantial uncertainties in their physical properties and distances. For these reasons, we do not conduct detailed hadronic fitting in this work, leaving more complex models until further multiwavelength data is collected.

\subsection{The Candidate PWN in the Galactic Environment}
\label{sec:GalEnv}

For our modeling, we fixed the distance to the primary XMM-Newton source to be $d = 1 \:\rm{kpc}$, with its galactic coordinates ($l = 147.0^\circ$, $b = -1.8^\circ$) giving a position in the outskirts of the galactic plane. Supposing that the candidate PWN is of age $\lesssim 10 \:\rm{kyr}$, typical of $> \rm{TeV}$ PWN \citep{PWNhess}, it is likely that this system is part of one of the well-mapped spiral arms opposite the galactic center, the homes of star formation in the more rural outer regions of the Milky Way. For comparison, we draw on the work by \cite{MWsfrMap}, who mapped in detail the star formation regions of the outer arms of our side of the Milky Way, a proxy map of environments where the production of a PWN is likely.

\begin{figure}
    \centering
    \includegraphics[width=\columnwidth]{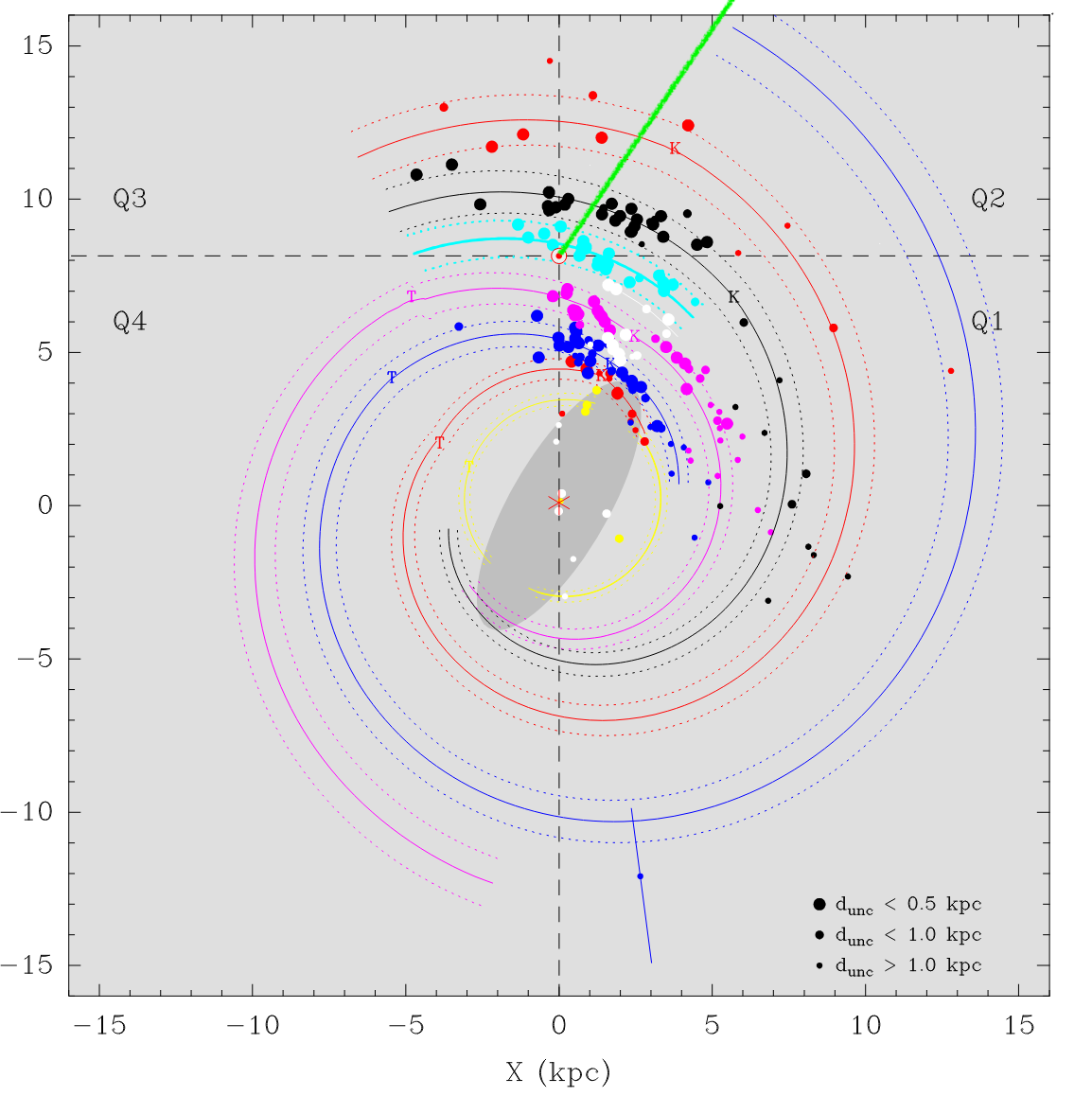}
    \caption{Reproducing Figure 10 from \cite{MWsfrMap}, a top-down view of the arms and star-forming regions in the Milky Way galaxy. Given the direction of $l=147.0^\circ$ towards our primary source (green line), it is likely that this candidate PWN is a member of the local (cyan, $d < 1 \:\rm{kpc}$), Perseus (black, $d < 2\:\rm{kpc}$), or perhaps Norma-Outer (red, $d < 6 \:\rm{kpc}$) spiral arms.}
    \label{fig:GalaxyPosition}
\end{figure}

Reproducing Figure 10 from \cite{MWsfrMap} and including the vector along $l = 147.0^\circ$ for our primary source, we can show in Figure \ref{fig:GalaxyPosition} that the distance to our primary source is likely $\lesssim 2 \:\rm{kpc}$ given its likely membership in the local arm or Perseus arm, though a distance of $\approx 4 \:\rm{kpc}$ would be possible if the system is a member of the Norma-Outer arm. 1LHAASO J0343+5254u is an attractive target for further observations, as it is likely closer to us than many other galactic UHE sources and occupies a less busy region of the galactic plane.

\subsection{Radio Context}
\label{sec:Radio}

\begin{figure}
    \centering
    \includegraphics[width=\columnwidth]{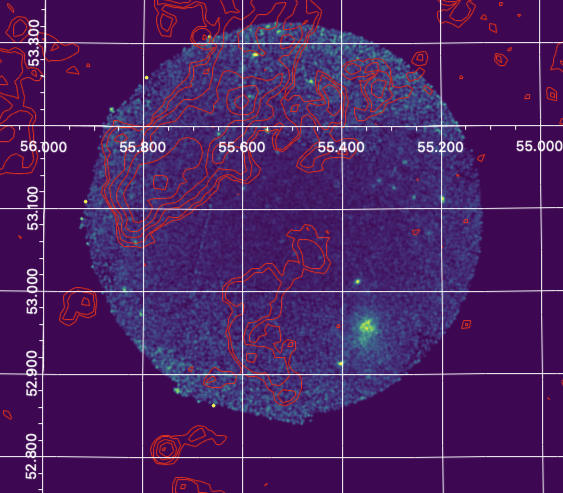}
    \caption{
    The XMM-Newton image in $0.2-12 \:\rm{keV}$ keV, overlaid with contours of $^{12}$CO ($J$=1--0) line emissions within $ -40 \:\rm{km/s} < V_{\rm LSR} < 10 \:\rm{km/s}$ obtained by the Nobeyama radio telescope.
    }
    \label{fig:Radio}
\end{figure}

We have been conducting observations of radio line emissions using the Nobeyama Radio Observatory (NRO) 45-m radio telescope to search for associated molecular clouds in the 1LHAASO J0343+5254u region (Tsuji et al. in prep). Five molecular clouds are detected within the gamma ray regions of the LHAASO sources; four have derived distances and $n_H$ in the range $0.3-1 \:\rm{kpc}$ and $200-800 \:\rm{cm^{-3}}$, with the fifth having $d = 4.0 \pm 1.0 \:\rm{kpc}$ and $n_H = 2 \times 10^3 \:\rm{cm^{-3}}$ (related to a nearby optical/IR asymptotic giant star). These radio sources may be molecular clouds where outflows of energetic protons from the PWN interact with ambient protons to eventually produce gamma rays via neutral pion decay; however, none of them have substantial X-ray flux in our XMM observations, shown in Figure \ref{fig:Radio}. There is also no apparent radio counterpart at the position of the primary X-ray source, nor any extended radio emission pointing to any of the other high-energy sources in this region and relating a kicked pulsar within the candidate PWN back to a radio relic at its birth position.

Upper limits limit for radio continuum flux at the primary X-ray position are $1.2 \times 10^{-15} \:\rm{erg/s/cm^2}$ at $408 \:\rm{MHz}$ and $4.3 \times 10^{-15} \:\rm{erg/s/cm^2}$ $1420 \:\rm{MHz}$. These limits are not particularly constraining, exceeding the fluxes predicted by the leptonic model by several orders of magnitude.

The X-ray emission of our candidate PWN could feasibly be located at the same distance of the detected molecular clouds. Particularly, a partial shell-like structure, which was mentioned in \cite{J0343orig}, is visible to the east of the primary X-ray source, and the X-ray source seems to be located in the vicinity of the half-shell-like molecular cloud. If they are related to each other, the distance to the primary X-ray source would be $\lesssim0.3 \:\rm{kpc}$, within the local spiral arm.

\section{Conclusions and Next Steps}
\label{sec:Conc}

We have reported the discovery of a new candidate PWN in the outer regions of the galactic plane, which may be linked to 1LHAASO J0343+5254u. With its overall featureless X-ray spectrum ($\Gamma_X = 1.9 \pm 0.1$), extended morphology, and spatially varying X-ray hardness, the source is similar to other PWN related to VHE and UHE emission, like the Eel or Boomerang nebulae.  We do not detect a radio point source or substantial X-ray pulsations at this primary source, but further investigations may detect these concrete signs of PWN classification.

We construct an SED linking the primary XMM-Newton source to observed LHAASO emission, using the combined X-ray and VHE emission, plus upper limits established by \textit{Swift}-BAT, \textit{Fermi}-LAT, and VERITAS nondetection. We establish a leptonic (synchrotron + IC) model for this combined emission, finding that the leptonic model can explain the MWL emission by invoking elevated ambient IR fields for IC emission. If the IR fields in the environment around our PWN are not high enough to produce all the observed LHAASO emission, then the difference may be made up of gamma rays from hadronic processes in nearby molecular clouds.

In this leptonic model, our primary source may be a VHE PWN similar to other systems like the Eel \citep{EelPWN}, the Boomerang \citep{BoomerangPWN}, HESS J1825-137 \citep{HESSpwn}, or the Dragonfly \citep{DragonflyPWN}.

\subsection{Next Steps}

Additional X-ray observations above $10 \:\rm{keV}$ could much more conclusively constrain the magnetic field of the PWN and the total synchrotron emissivity in the PWN. Measuring the X-ray cutoff of the power law spectrum of the candidate PWN will allow for direct constraints on these physical parameters, so we will propose observations with the Nuclear Spectroscopic Telescope Array \citep{NuStar} to directly test the model proposed herein and much more tightly constrain $B$ in the candidate PWN.

Particularly high-resolution X-ray observations with Chandra could resolve intricate details of the candidate PWN, including X-ray structure in the central $1.8 \arcmin$ or an X-ray point source in the central region. The detection of a specific X-ray point source would allow for substantially more focused followup at a presumptive pulsar, which is not resolved in our XMM-Newton observations.

Finally, radio searches in the region of the primary XMM-Newton source could identify the position of the pulsar itself, allowing for much more detailed physical modeling of an evolving PWN system in a multiwavelength context. In other PWN systems, the radio pulsar is typically located within the X-ray region, suggesting that a radio search for a pulsar within our candidate PWN could be limited to the region immediately around the candidate PWN. The discovery of a radio pulsar would cement PWN classification and allow for much more detailed evolutionary modeling of the X-ray emission region.

There are several astronomical unknowns that inhibit a detailed and tightly constrained hadronic model. First, lacking a evolutionary model for a pulsar in the heart of the candidate PWN prevents physical conclusions about the flow of particles into a target molecular cloud, the first step in a hadronic model. Speaking of, there is substantial uncertainty about the molecular cloud that would serve as a target. Finally, the 2021 spectrum for LHAASO J0343+5254u may be updated with more recent LHAASO data and/or detections with other gamma ray observatories. For these reasons, we do not conduct detailed hadronic fitting in this work, but the additional of further MWL data may prove fruitful for more complicated lepto-hadronic modeling of the system.

While the candidate PWN is a feasible counterpart to the LHAASO emission, the region around 1LHAASO J0343+5254u may have a nuanced spatial geometry; the two KM2A sources identified in \cite{1LHAASO} in particular may mean that the VHE emission in this region is possibly two independent confused or unresolved systems. Continuing analysis of the nearby \textit{Fermi}-LAT gamma ray source and radio observations of molecular clouds in the region will constrain the MWL landscape near 1LHAASO J0343+5254u. In-progress, detailed examinations with other gamma ray telescopes like HAWC and VERITAS will also constrain the environment around 1LHAASO J0343+5254u.

\software{FTools \citep{FTools}, Xspec \citep{Xspec}, DS9 \citep{DS9}, FermiPy \citep{FermiPy}, XMMSAS \citep{XMMSAS}}

\acknowledgments

This research has made use of data and/or software provided by the High Energy Astrophysics Science Archive Research Center (HEASARC), which is a service of the Astrophysics Science Division at NASA/GSFC. We are grateful for Alison Mitchell for substantial guidance and comments on this work. Support for K. Mori and J. Woo was provided by NASA through the XMMNC22 grant.

\bibliography{main}{}

\end{document}